\documentclass[preprint2]{aastex6}
\usepackage{epsfig}
\usepackage{amsfonts,amssymb,amsmath}

\begin{document}

\title{Constraints on the Evolution of the Galaxy Stellar Mass Function II: Quenching TimeScale of Galaxies and its Implication for their Star
Formation Rate}
\author{E. Contini$^{1,2}$, X. Kang$^{1}$, A.D. Romeo$^{1}$, Q. Xia$^{1}$, S.K. Yi$^{2}$}

\affil{$^1$Purple Mountain Observatory, the Partner Group of MPI f\"ur Astronomie, 2 West Beijing Road, Nanjing 210008, China}
\affil{$^2$Department of Astronomy and Yonsei University Observatory, Yonsei University, Yonsei-ro 50, Seoul 03722, Republic of Korea}

\email{contini@pmo.ac.cn}

\email{kangxi@pmo.ac.cn}

\email{yi@yonsei.ac.kr}

\begin{abstract} 
We study the connection between the observed star formation rate-stellar mass (SFR-$M_*$) relation and the evolution of 
the stellar mass function (SMF) by means of a subhalo abundance matching technique coupled to merger trees extracted 
from a N-body simulation. Our approach consist of forcing the model to match the observed SMF at redshift $z \sim 2.3$, 
and let it evolve down to $z \sim 0.3$ according to a $\tau$ model, an exponentially declining functional form which 
describes the star formation rate decay of both satellite and central galaxies. 
In this study, we use three different sets of SMFs: ZFOURGE data from Tomczak et al.; UltraVISTA data from Ilbert et al. and 
COSMOS data from Davidzon et al. We also build a mock survey combining UltraVISTA with ZFOURGE. 
Our modelling of quenching timescales is consistent with the evolution of the SMF down to  
$z \sim 0.3$, with different accuracy depending on the particular survey used for calibration. We tested our model against 
the observed SMFs at low redshift and it predicts residuals (observation versus model) within $1\sigma$ observed scatter along most 
of the stellar mass range investigated, and with mean residuals below 0.1 dex in the range $\sim [10^{8.7}-10^{11.7}] M_{\odot}$.
We then compare the SFR-$M_*$ relation predicted by the model with the observed one at different redshifts. The predicted SFR-$M_*$ 
relation underpredicts the median SFR at fixed stellar mass relative to observations at all redshifts. 
Nevertheless, the shapes are consistent with the observed relations up to intermediate-mass galaxies, followed by a rapid decline 
for massive galaxies.

\end{abstract}


\section[]{Introduction} 
\label{sec:intro}

The stellar mass function (SMF) is a fundamental statistic in astrophysics that helps us to improve our knowledge 
of galaxy formation and evolution. The SMF directly describes the distribution of the galaxy stellar mass, and the 
study of its evolution as a function of time can provide important information on the star formation history of 
galaxies. 

As pointed out by many authors recently (e.g. \citealt{leja15,tomczak16,contini17}), a peculiar aspect of 
the evolution of the SMF is not yet clear, that is, its relation with the star formation rate-stellar mass (SFR-$M_*$) 
relation (e.g., \citealt{tomczak16,contini17} and references therein). In fact, 
it is in principle possible to connect the observed SMF at any redshift with the SFR-$M_*$ relation to obtain 
the evolution of the SMF, assuming that all processes (not only star formation) responsible for the growth of galaxies 
are considered. A similar approach has been followed by several authors (e.g., \citealt{conroy09,leja15,tomczak16,contini17}) 
and all these studies concluded that, considering reasonable assumptions for mergers of galaxies and stripping of stars,
which are the most important physical processes that act on the galaxy's stellar mass, there still are some
discrepancies between the observed and modeled SMFs. Nevertheless, it appears now clear the need for a mass-dependent slope 
for the SFR-$M_*$ relation in order to reconcile the observed evolution of the SMF with that inferred by connecting 
the SFR-$M_*$ relation with the observed SMF at high redshift (\citealt{contini17} and references therein). In 
accordance with these studies (e.g., \citealt{leja15,tomczak16,contini17}), 
neither mergers or stellar stripping altogether, nor hidden low-mass quiescent galaxies not detected are likely to be 
responsible for the mismatch (which appears to be in the range 0.2-0.5 dex over the whole stellar mass range
investigated). 

The question arises quite spontaneously: What is responsible for the mismatch? There might be several reasons, 
ranging from possible physical processes that we have not understood correctly, to the correct shape of the SFR-$M_*$
relation and the scatter around it, or even to uncertainties in the observed SMF at high redshift. Answering  
this question is not an easy task simply because, in following the above-mentioned approach, many uncertainties 
are mutually connected. Taking advantage of the ZFOURGE survey, \cite{tomczak16} generated star-fomation histories (SFHs)
of galaxies by means of the SFR-$M_*$ relation redshift dependent, and integrated the set of SFHs with 
time to obtain mass-growth histories to compare to those inferred from the evolution of the SMF of \cite{tomczak14}. 
Their result, despite the reasonable agreement between the observed and inferred SMFs, suggests that either the SFR 
measurements are overestimated, or the growth of the \cite{tomczak14} mass function is too slow, or even both.

In \cite{contini17} (hereafter PapI) we followed an approach similar to that described above. In order to match 
the SMF at high redshift and predict its evolution with time, we used an analytic model which considers mergers and 
stripping coupled with an abundance-matching technique to set the galaxy stellar mass.
The main goal in that study was to reconcile the evolution of the SMF coupled to the observed SFR-$M_*$ relation, 
by including real merger trees from N-body simulations that provide the accretion history of galaxies and a
prescription for stellar stripping. We concluded that the SFR-$M_*$ relation must be mass and redshift dependent,
and both mergers and stellar stripping are important processes for the shape of the massive end of the SMF. 
Moreover, by testing two different sets of SMFs coupled to the same SFR-$M_*$ relation 
and the same modeling for mergers and stellar stripping, we found different evolutions down to low redshift.
As a consequence, either the observed SMF at high-z bears uncertainties too large to be conclusive (as we concluded in PapI), 
or the growths of the SMF described by different observations are not self-consistent.

In this study, we want to address this topic by following an approach slightly different from the one used 
in PapI. The core of the method remains unchanged, meaning that we use merger trees constructed from the same 
N-body simulation and the same prescription for stellar stripping, and we set up the stellar mass of galaxies by using an abundance-matching 
technique. In order to describe the star formation histories of galaxies, we match the observed SMF at high redshift 
and assign an SFR by means of the observed SFR-$M_*$ relation (as done in PapI), but we let the SFR decay exponentially with time 
according to a quenching timescale that is set for each galaxy. \footnote{In PapI we assign SFRs to galaxies according to their 
mass at each time step.}
The goals are to:
\begin{itemize}
 \item [1)] calibrate our model with the observed SMF from different surveys and study their evolution,
 \item [2)] reduce the mismatch between the observed and inferred SMFs at low redshift, and
 \item [3)] compare the predicted SFR-$M_*$ relation with the observed one as a function of time.
\end{itemize}

The paper is structured as follows. In Section \ref{sec:method} we describe in detail our approach, the method followed, and 
modeling, and we give a brief introduction to the surveys considered. In Section \ref{sec:results} we show our results, which 
will be fully discussed in Section \ref{sec:discussion}. Finally, in Section \ref{sec:conclusions} we give our conclusions.
Throughout this paper we use the standard cosmology summarized in Section \ref{sec:method}. The stellar masses are given in 
units of $M_{\odot}$, and we assume a \cite{chabrier03} IMF.

\section[]{Methods}  
\label{sec:method}

The simulation used in this paper is based on \cite{kang12} and fully described in PapI. We refer the readers to those papers for the 
details of the simulation, while here we give an overview of our approach.

Similarly to the approach used in PapI, we use the so-called subhalo abundance matching (ShAM) technique to populate dark matter haloes with galaxies
(\citealt{vale04}), which is now widely used in numerical simulations to connect galaxies with dark matter structures
(\citealt{guo14,yamamoto15,chaves-montero16} just to quote some recent works). For more details concerning the ShAM technique, we 
refer readers to PapI or the studies quoted above. 

In order to study the evolution of the SMF we need to connect galaxies with haloes at different times. Despite the outputs of the simulation 
being stored in 60 snapshots, from $z \sim 100$ to the present time, we start by matching the observed SMF at $z_{match}=2.3$,
so that the SMF predicted by our model is exactly the observed one. The algorithm reads the merger trees constructed from our simulation and links galaxies 
and dark matter haloes in a one-to-one correlation by firstly sorting them in mass. As in PapI, since new haloes can form after $z_{match}$, we populate 
them with galaxies by using the stellar mass-halo mass relation at that redshift.

After $z_{match}$ we let galaxies evolve according to their merger histories (given by the merger trees) and to their star formation histories. At 
$z_{match}$ (or the redshift when they are born if it happens after $z_{match})$ we assign SFRs to galaxies by means of the SFR-$M_*$ relation observed 
at that redshift, and we let the SFRs evolve down to $z \sim 0.3$ according to a $\tau$ model that describes the star formation histories (SFHs) of both satellite 
and central galaxies (see Section \ref{sec:model}). 

Galaxies can lose a fraction of their stellar mass once they become satellites via stellar stripping, due to gravitational interactions with their hosts. Our 
model considers stellar stripping, and here we summarize the prescription used for this process. For further details, we refer the reader to PapI.

Following \cite{contini14}, who have analyzed the channels for the formation of the intracluster light, we model stellar stripping in a very similar fashion, that is,
by assuming an exponential form for the stellar mass lost:
\begin{equation}
 M_{lost}^{z=z_i} = M_{*(s1,s2)}^{z=z_i} \cdot \left( 1-\exp \left(\frac{-\eta (M_{halo})(t_{infall}-t_i)}{\tau_{s1,s2}}\right) \right) \, ,
\end{equation}
where $t_{infall}$ is the lookback time when the galaxy last entered a cluster (i.e., became a satellite), $t_i$ is the lookback  time at $z=z_i$, and $\tau_{s1,s2}$ 
are two normalizations arbitrarily set to 30 Gyr and 15 Gyr, respectively, for satellites $s_1$ associated with a distinct dark matter substructure and orphan 
galaxies $s_2$ (galaxies not associated with any dark matter substructure). We set $\tau_{s1} > \tau_{s2}$ in order to consider the effect of dark matter in 
shielding the galaxy from tidal forces (see PapI for further details).
The term $\eta (M_{halo})$ is the stripping efficiency, and it provides the strength of stellar stripping as a function of the main halo mass $M_{200}$,
which is the mass within the virial radius $R_{200}$.

In \cite{contini14} we considered a second important process for the formation of the intracluster light, i.e. the so-called "merger channel" (see also  
\citealt{giuseppe}). We simply assume that when two galaxies merge, 30\% of the satellite stellar mass becomes unbound and goes to the diffuse component. A 
similar prescription has been used also in other semi-analytic models (see, e.g. \citealt{pigi07}).

\subsection[]{Model of star formation}  
\label{sec:model}
In this section, we describe in detail how we model the galaxy stellar mass growth from high to low redshift due to star formation. As explained above where 
we described our approach, we force the algorithm to match the observed SMF at redshift $z_{match}=2.3$. In PapI we assigned SFRs to galaxies making use of 
the $SFR-M_*$ relations observed by \cite{tomczak16}, which means that each galaxy was assigned an SFR according to its stellar mass \emph{and redshift}. In this 
study, we model the SFHs of galaxies in a different way: 

\begin{itemize}
\item[1)] We assign an SFR to each galaxy according to the $SFR-M_*$ relation either at $z=z_{match}$ or, if a galaxy forms after $ z_{match}$, 
          at $ z=z_{form}$.
\item[2)] After $ z_{match}$ or $ z_{form}$, the SFR of each galaxy will evolve according to functional forms that take into account 
	  information such as type (central or satellite), stellar mass, and, more importantly, a quenching timescale.
\end{itemize}
This approach is clearly different from the one adopted in PapI and adds more direct information about the time galaxies spend before being quenched. 
We treat the SFHs of central and satellite galaxies separately. For centrals, we use a prescription very similar to the one adopted in \cite{noeske07}, with 
some differences:

\begin{equation} \label{eq:sfrcen}
SFR_{cen}(t)=SFR_{match/form} \cdot \exp{\left(-\frac{t}{\tau_c}\right)} \, ,
\end{equation}
where $\tau_c$ is the quenching timescale. $\tau_c$ is derived from the following equation:

\begin{equation} \label{eq:taus}
\tau_{c,s} = 10^{11.7} \cdot \left(\frac{M_*}{M_{\odot}} \right)^{-1} \cdot (1+z)^{-1.5} \; [Gyr],
\end{equation}
where $M_*$ is the stellar mass at $z=z_{match/form}$ and a random scatter in the range $\pm 20 \%$ is assigned as a perturbation. Our prescription is different from 
the original one (\citealt{noeske07}) in the sense that we consider only the stellar mass (rather than the baryonic mass) and add a redshift-dependent correction. We will 
come back to the reason for this choice later.

To model the SFHs of satellites, we take advantage of a revised version of the so-called \emph{delayed-then-rapid} quenching mode suggested by \cite{wetzel13}, 
in which the SFRs of satellite galaxies evolve like those of centrals for $2-4$ Gyr after infall, and then quench rapidly. For satellites, we assume a delayed quenching, 
after which they quench according to a quenching timescale $\tau_s$. The quenching timescale $\tau_s$ is set at $z=z_{match/form}$ and is given by equation \ref{eq:taus} 
if a galaxy becomes satellite before $z_{match}$, and it is set as a random fraction between 0.1 and 0.5 of $\tau_c$ for the other galaxies. These choices altogether guarantee that 
$\tau_s < \tau_c$ for every galaxy. Hence, the SFR of satellites evolves as described by equation \ref{eq:sfrcen} if $$t_{since \, infall} < t_{delay} \, ,$$ where 
$t_{delay}$ is randomly chosen in the range [2-4] Gyr, and as

\begin{equation} \label{eq:sfrsat}
SFR_{sat}(t)=SFR_{match/form} \cdot \exp{\left(-\frac{t}{\tau_s}\right)} 
\end{equation}
thereafter. The parameter $SFR_{match/form}$ in equations \ref{eq:sfrcen} and \ref{eq:sfrsat} is set at $z=z_{match/form}$ and derived as done 
in PapI, by following Equation 2 in \cite{tomczak16}:
\begin{equation}\label{eq:tomc_par}
  \log(SFR \, [M_{\odot}/yr]) = s_0 -\log \left[ 1+\left( \frac{M_*}{M_0}\right)^{\gamma} \right]  \, ,
\end{equation}
where $s_0$ and $M_0$ are in units of $\log(M_{\odot}/yr)$ and $M_{\odot}$ respectively. In Equation \ref{eq:tomc_par} the model considers a random scatter in the 
range $\pm 0.2$ dex as a perturbation, as suggested by A. Tomczak (2017, private communication). \cite{tomczak16} find that such a parameterization works well even if quiescent 
galaxies are considered. Then, as in PapI, where we showed that this parameterization agrees fairly well with the evolution of the SMF with time, we assume Equation \ref{eq:tomc_par} 
to be valid for all galaxies. Hence, $s_0$ and $M_0$ are given by (Equation 3 in \citealt{tomczak16})
\begin{displaymath}
s_0 = 0.195  +  1.157 z  -  0.143 z^2
\end{displaymath}
\begin{equation}
\label{eq:parameterisation_all}
 \log(M_0 )= 9.244  +  0.753 z  -  0.090 z^2  
\end{equation}
\begin{displaymath}
 \gamma = -1.118
\end{displaymath}
Equation \ref{eq:tomc_par} and the set of equations \ref{eq:parameterisation_all} describe the evolution with time of the SFR-$M_*$ relation with a mass-dependent slope, which 
has been demonstrated to be necessary in order to reproduce the evolution of the SMF with time (see, e.g., \citealt{leja15,tomczak16,contini17}).

In this study, we have chosen to add a redshift term in the equation that provides the quenching timescales of both satellites and centrals. In fact, it has 
been proved that the quenching timescale, at least for satellite galaxies, depends on the redshift (see, e.g. \citealt{tinker_wetzel10,wetzel13}) being shorter for galaxies 
accreted at higher redshift (\citealt{contini16}). Galaxies accreted at higher redshift tend to have a gas fraction $M_{cold}/M_*$ larger than those accreted later, to be less 
massive, and to reside in lower halo mass and eject mass more efficiently (\citealt{contini16}). In principle, there is no justification to use the same assumption 
also for central galaxies. Nevertheless, it must be noted that in the original version of the model we applied this correction only to satellites. The plots we will show in Section 
\ref{sec:results} are the result of the model described above, which remarkably improved over the original version (no redshift-dependent quenching timescale for centrals).

Our model for star formation quenching described in this section considers both environmental and mass quenching. The first one is explicitly included in equations \ref{eq:sfrcen}
and \ref{eq:sfrsat}, the quenching being much faster for satellite galaxies. A sort of mass quenching (which is not that described in \citealt{peng10}) is instead implicit in the calculation 
of the quenching timescales, in a way that, for both satellite and central galaxies, the quenching is faster with increasing stellar mass and redshift. We will come back to this topic in 
Section \ref{sec:discussion}.

\subsection[]{Surveys}  
\label{sec:surveys}

In this study, we take advantage of observed high-redshift data from three different surveys: ZFOURGE (\citealt{tomczak14,straatman16}),
UltraVISTA (\citealt{ilbert13}) and COSMOS (\citealt{davidzon17}). These surveys differs from one another in the total area covered,
mass completeness limit, and number statistics. In this section, we want to summarize the main aspects of each survey, and we refer the 
reader to the relevant studies quoted above for further details.

The first set of data we considered in order to calibrate our model at $z=z_{match}$ and compare model predictions with observations at lower 
redshift has been constructed by \cite{tomczak14}. These authors used observations from the FourStar Galaxy Evolution Survey (ZFOURGE), 
which is composed of three $11' \times 11'$ pointings with coverage in the CDFS (\citealt{giacconi02}), COSMOS (\citealt{capak07}) and UKIDSS 
(\citealt{lawrence07}). 
The ZFOURGE fields also use HST imaging taken as part of the CANDELS survey (\citealt{grogin11,koekemoer11}), and in \cite{tomczak14}, the 
survey has been supplemented with the NEWFIRM Medium-Band Survey (\citealt{whitaker11}).

The second set of data (from \citealt{ilbert13}) is a merged catalog of data from different studies. The main branch comes from a photometric 
catalog of near-IR data from the UltraVISTA project (\citealt{mccracken12}) and optical data from the COSMOS project (\citealt{capak07}). COSMOS 
is a wide survey that covers $2 deg^2$ area and with multiwavelength (more than 35 bands) data.The UltraVISTA DR1 data release covers $1.5 deg^2$
in four near-IR filters Y, J, H, and $K_s$ and provides deeper data than the previous COSMOS near-IR datasets from \cite{mccracken10}. The major 
features of the set of data used by \cite{ilbert13} rely upon the large number of spectra available and, even more importantly, on the robustness 
of the photometric redshifts derived. 

\begin{figure*} 
\begin{center}
\includegraphics[scale=.95]{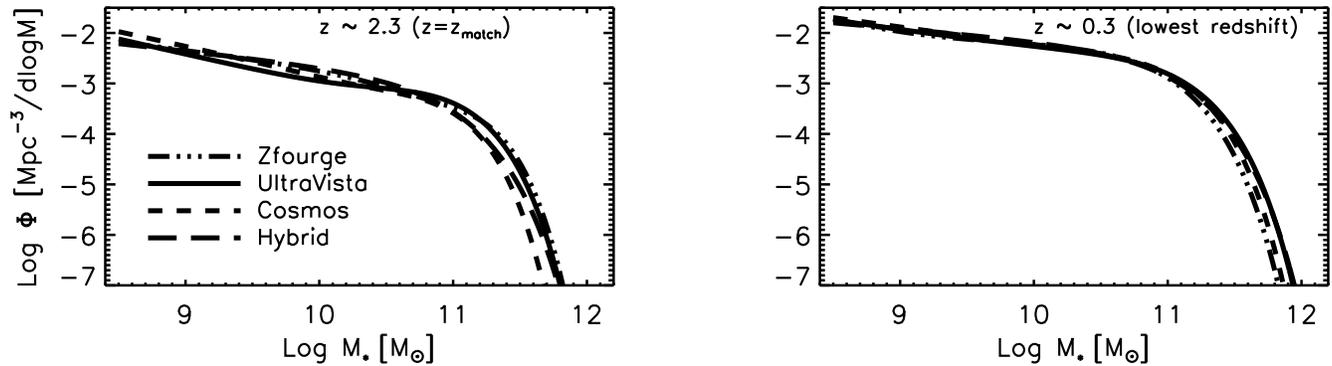} 
\caption{SMFs shown by different surveys, at redshift $z \sim 2.3$ (left panel) and $z \sim 0.3$ (right panel). Although they look very different 
at $z=z_{match}$, at low redshift the three surveys and our hybrid combination agree very well.}
\label{fig:surveys}
\end{center}
\end{figure*}

The third and last set of data considered comes from the COSMOS2015 catalog (\citealt{laigle16}), which includes ultradeep photometry from UltraVISTA 
(\citealt{mccracken12}), SPLASH (\citealt{capak12}), and Subaru/Hyper Suprime-Cam (\citealt{miyazaki12}). The COSMOS2015 catalog contains precise 
photometric redshifts and stellar masses for more than half a million objects over the 2$deg^2$ COSMOS field, and the deepest regions reach a 90\% 
completeness limit of $10^{10} M_{\odot}$ at $z=4$. It has been shown to provide a large-number statistics due to the large volume probed and improved 
depth compared with the previous versions of the catalog.

Due to the cross-match of different surveys in the catalogs described above, for the sake of simplicity and in order to avoid confusion, we hereafter 
call ZFOURGE the first catalog, UltraVISTA the second, and COSMOS the last. In Figure \ref{fig:surveys} we show the SMFs as obtained from 
these catalogs, at $z \sim 2.3$ (which is our $z_{match}$) in the left panel and at $z \sim 0.3$ in the right panel. At high redshifts, the three SMFs 
look different, and not only in the high-mass end where uncertainties are typically large, but also in the stellar mass range $[10^{9.5}-10^{10.5}] M_{\odot}$ with residuals 
that reach around $\sim 0.3-0.4$ dex. Despite that, the SMFs agree very well at $z \sim 0.3$ up to $10^{11.2-11.3} M_{\odot}$. The difference between the surveys at 
high redshift is mainly due to the mass completeness limit, which affects the fit of data. The completeness limits at $z \sim 2.3$ for the surveys we have 
chosen are (in $\log M_{\odot}$) 9.0 (ZFOURGE), 10.0 (UltraVISTA), and 9.3 (COSMOS). 

In order to overcome the problem of the mass completeness limit, we build another "mock survey" that takes advantage of the low-mass completeness limit
in ZFOURGE (which at odds does not sample the high-mass end) and UltraVISTA (which does). We derive the best-fit Schechter parameters by using a single 
Schechter function (\citealt{schechter76}) \footnote{For simplicity, we decided to use a single Schechter. Our choice is supported by the result found in 
\cite{tomczak14}, who showed that a single Schechter function is a good approximation of data at redshift $>2$.} defined as
\begin{equation}\label{eq:schechter}
 \frac{\Phi(M)dM}{\ln(10)} = \phi^* \left[10^{(M-M^* )(1+\alpha^* )}\right]\exp \left( -10^{(M-M^* )}\right) dM
\end{equation}
where $M = \log(M/M_{\odot})$, $\alpha^*$ is the slope at the low-mass end, $\phi^*$ is the normalization and $M^*$ is the characteristic mass. A mock realisation 
of data has been derived as follows. We have used the ZFOURGE fit to construct mock data from low mass up to $10^{11} M_{\odot}$ and the UltraVISTA fit for the rest of the stellar 
mass range. By using equation \ref{eq:schechter}, we have calculated the best-fit Schechter parameters for this set of mock data. The "mock survey" is named HYBRID, and its 
SMF is shown in Figure \ref{fig:surveys}. In Table \ref{tab:parameters} we list the sets of parameters used to calibrate the model at $z=z_{match}\simeq 2.3$ with our 
four surveys.

\begin{table}
\caption{Logarithm of the normalization, slope at the low-mass end, logarithm of the characteristic mass of the SMFs at $z=z_{match}$ and $\chi^2_{red}$ of the fit 
for the four surveys}
\begin{center}
\begin{tabular}{llllll}
\hline
Survey & $\log(\phi^* )$ & $\alpha^*$ & $\log(M^*)$ & $\chi^2_{red}$ \\
\hline
ZFOURGE     & $-3.59\pm0.14$  & $-1.43\pm0.08$ & $11.13\pm0.13$ & 0.30\\
UltraVISTA  & $-3.29\pm0.07$  & $-1.34\pm0.09$ & $10.79\pm0.02$ & 0.22 \\
COSMOS      & $-3.18\pm0.09$  & $-1.30\pm0.07$ & $10.80\pm0.02$ & 0.05 \\
HYBRID       & $-3.25\pm0.06$  & $-1.29\pm0.05$ & $10.84\pm0.02$ & 0.06 \\

\hline
\end{tabular}
\end{center}
\label{tab:parameters}
\end{table}

\section{Results}
\label{sec:results}

We calibrate our model with each of the surveys described above and let galaxies evolve down to $z\sim 0.3$, which is 
the lowest redshift for which we have observed data to be compared with model predictions. In this section, we study 
the evolution of the SMF down to $z\sim 0.3$ as predicted by each flavor of the model.

\begin{figure*} 
\begin{center}
\begin{tabular}{cc}
\includegraphics[scale=.95]{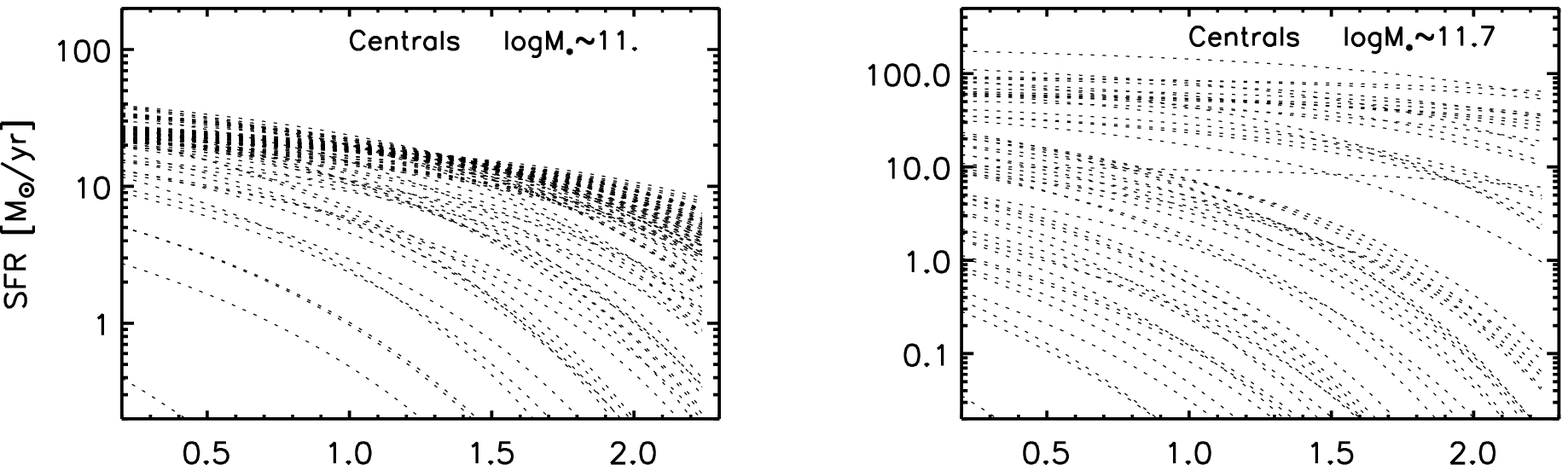} \\
\includegraphics[scale=.95]{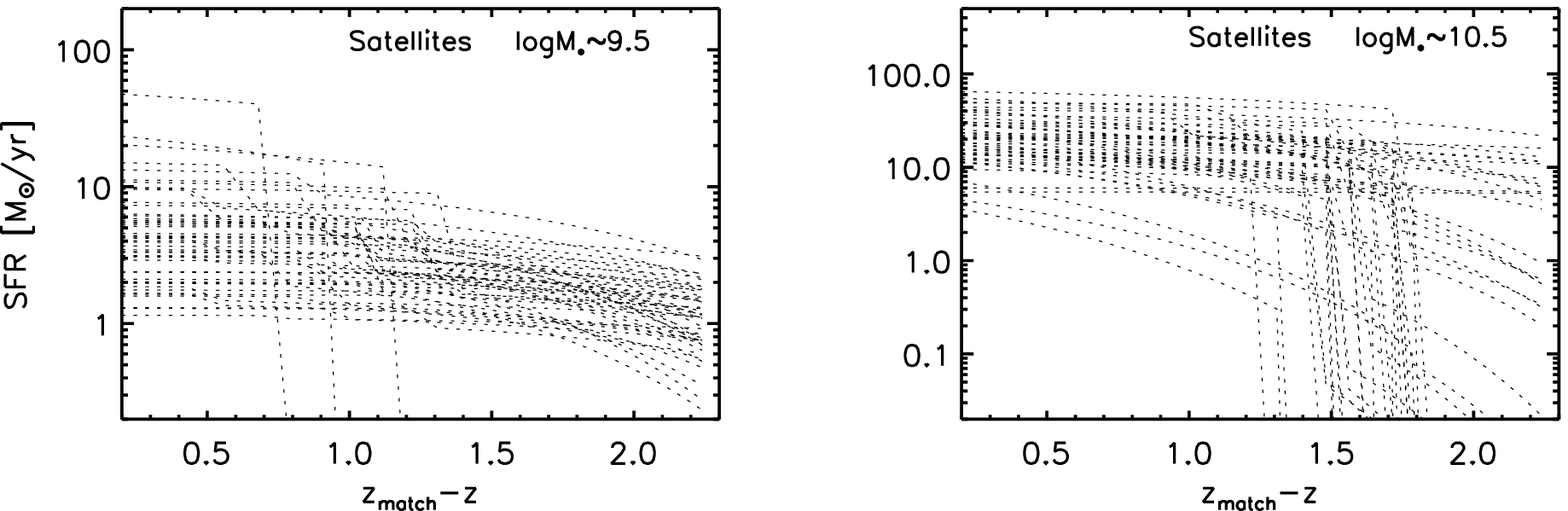} \\ 
\end{tabular}
\caption{Several examples of star formation rate histories for different kinds of galaxies: centrals with  $\log M_* \sim 11$ 
(top left panel), centrals with $\log M_* \sim 11.7$ (top right panel), satellites with $\log M_* \sim 9.5$ (bottom left panel) and 
satellites with $\log M_* \sim 10.5$ (bottom right panel). Both centrals and satellites have been chosen such that the full samples 
could show a wide variety of $\tau_c$ and $\tau_s$.}
\label{fig:sfrh}
\end{center}
\end{figure*}

The evolution of the SMF depends on our modeling described in Section \ref{sec:method}, and it is linked to the 
star formation quenching of galaxies. Figure \ref{fig:sfrh} shows several examples of star formation rate histories (SFRH) for central 
(upper panels) and satellite (bottom panels) galaxies, in different stellar mass ranges and for a large variety of quenching 
timescales $\tau_c$ and $\tau_s$. The SFRH of centrals is very regular, as shown by the exponential decline, and its 
amplitude clearly depends on the SFR at $z=z_{match,form}$ and $\tau_c$. On the other hand, the SFRH of satellites is
very diverse. A percentage of satellites have SFRHs very similar to central galaxies, and others show SFRHs with rapid declines
after infall or even several rapid declines. In the first case, they are satellites accreted in a short time after either $z=z_{match}$
or $z=z_{form}$, so that they have $\tau_c \sim \tau_s$ because the galaxy itself did not have much time to grow. In the 
other case, $\tau_s << \tau_c$, so the satellite galaxy after infall keeps forming stars as a central for a period of time in the 
range [2-4] Gyr and then its SFR declines quickly. In a few cases, a galaxy experiences several rapid declines because
it can spend some time as a satellite and then become central again.

\begin{figure*} 
\begin{center}
\includegraphics[scale=.9]{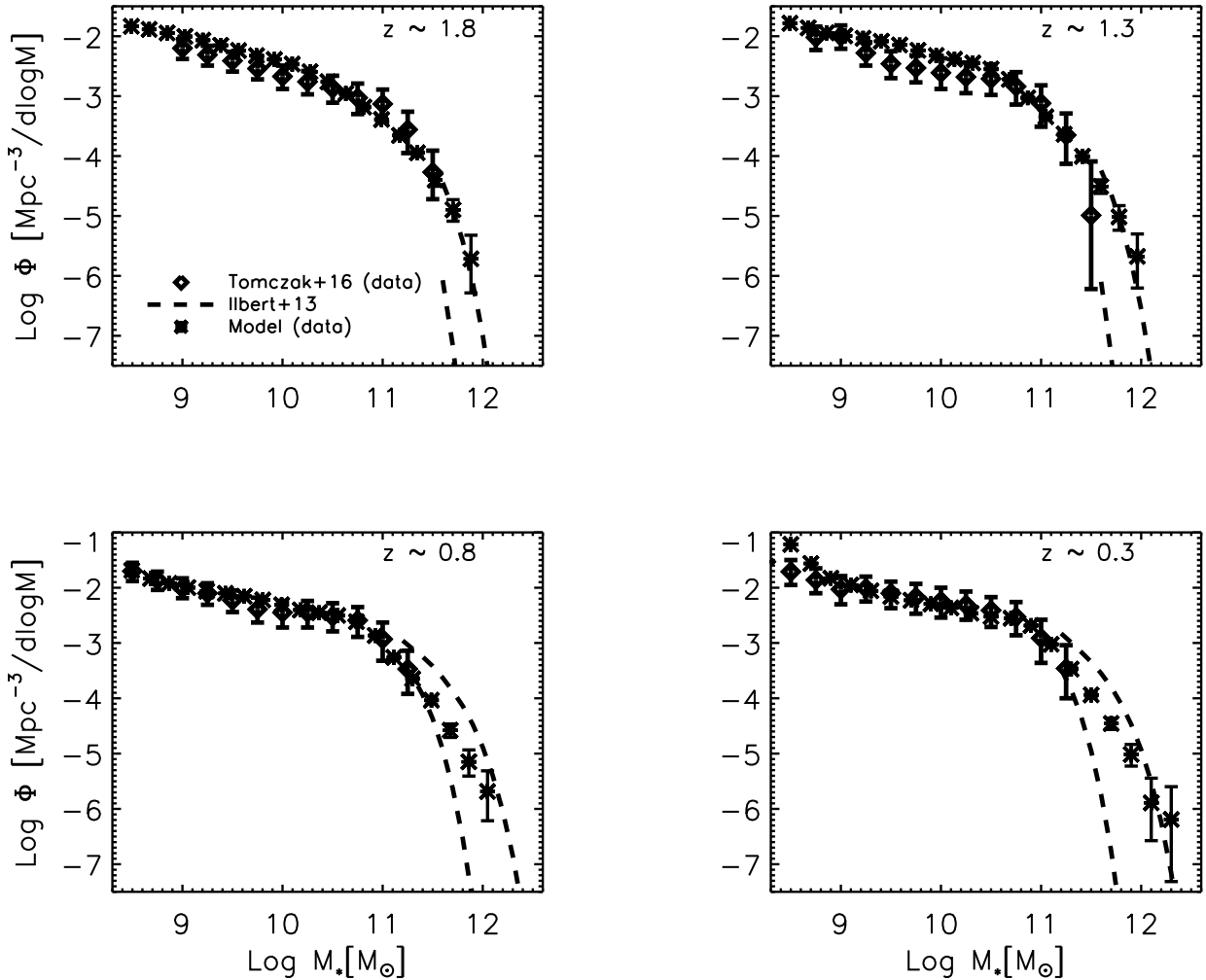} 
\caption{Evolution of the SMF from $z \sim 1.8$ down to $z \sim 0.3$ predicted by our model calibrated on the ZFOURGE survey at $z_{match}=2.3$. Stars and diamonds represent model and observed data, respectively,
while the dashed lines represent the \cite{ilbert13} fit of UltraVISTA data in the high-mass end (3$\sigma$ range).}
\label{fig:smf_zf}
\end{center}
\end{figure*}

\begin{figure*} 
\begin{center}
\includegraphics[scale=.9]{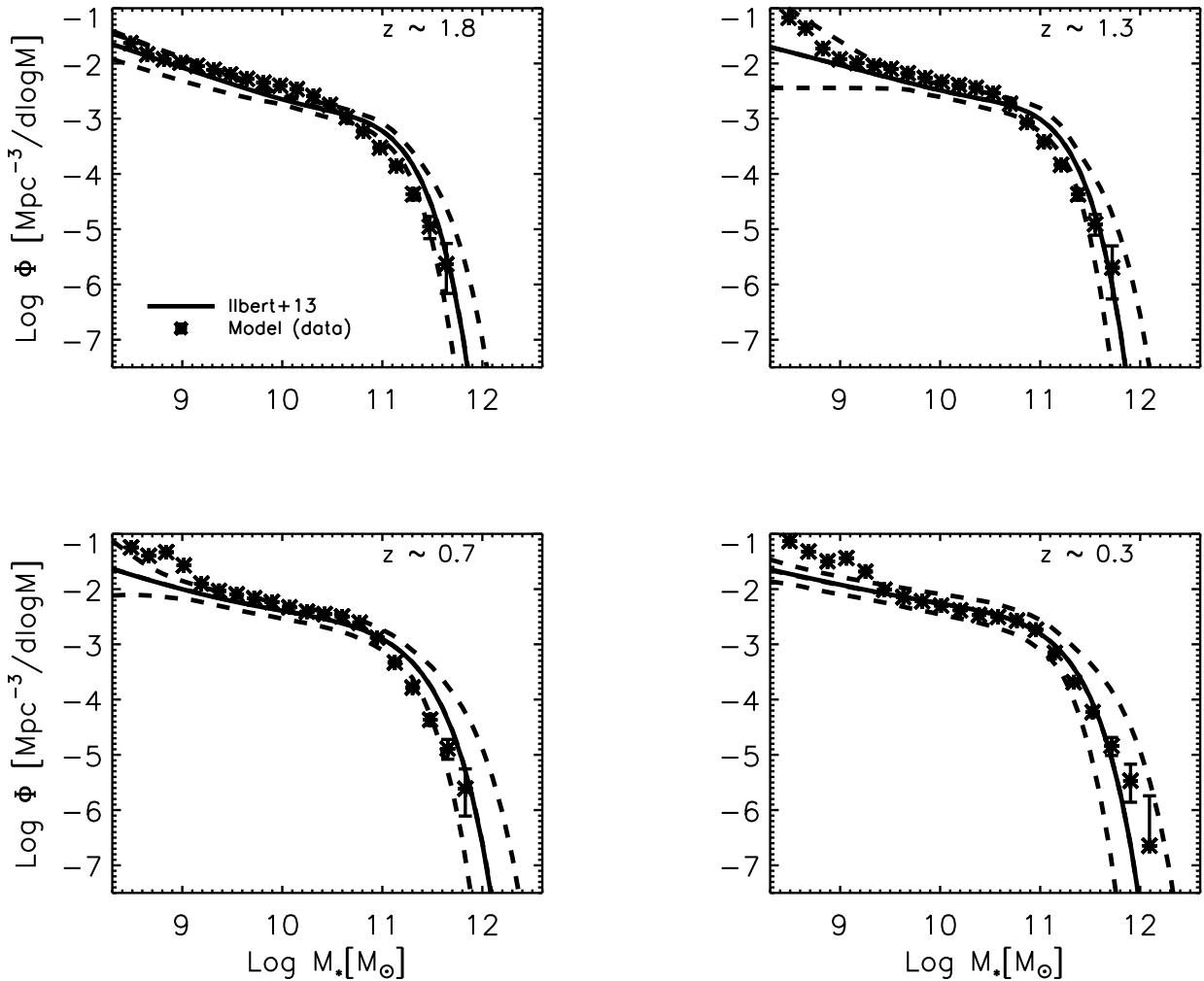} 
\caption{Evolution of the SMF from $z \sim 1.8$ down to $z \sim 0.3$ predicted by our model calibrated on the UltraVISTA survey at $z_{match}=2.3$. Stars represent model data,
while the dashed lines represent the \cite{ilbert13} fit of the UltraVISTA data (3$\sigma$ range).}
\label{fig:smf_uv}
\end{center}
\end{figure*}

\begin{figure*} 
\begin{center}
\includegraphics[scale=.9]{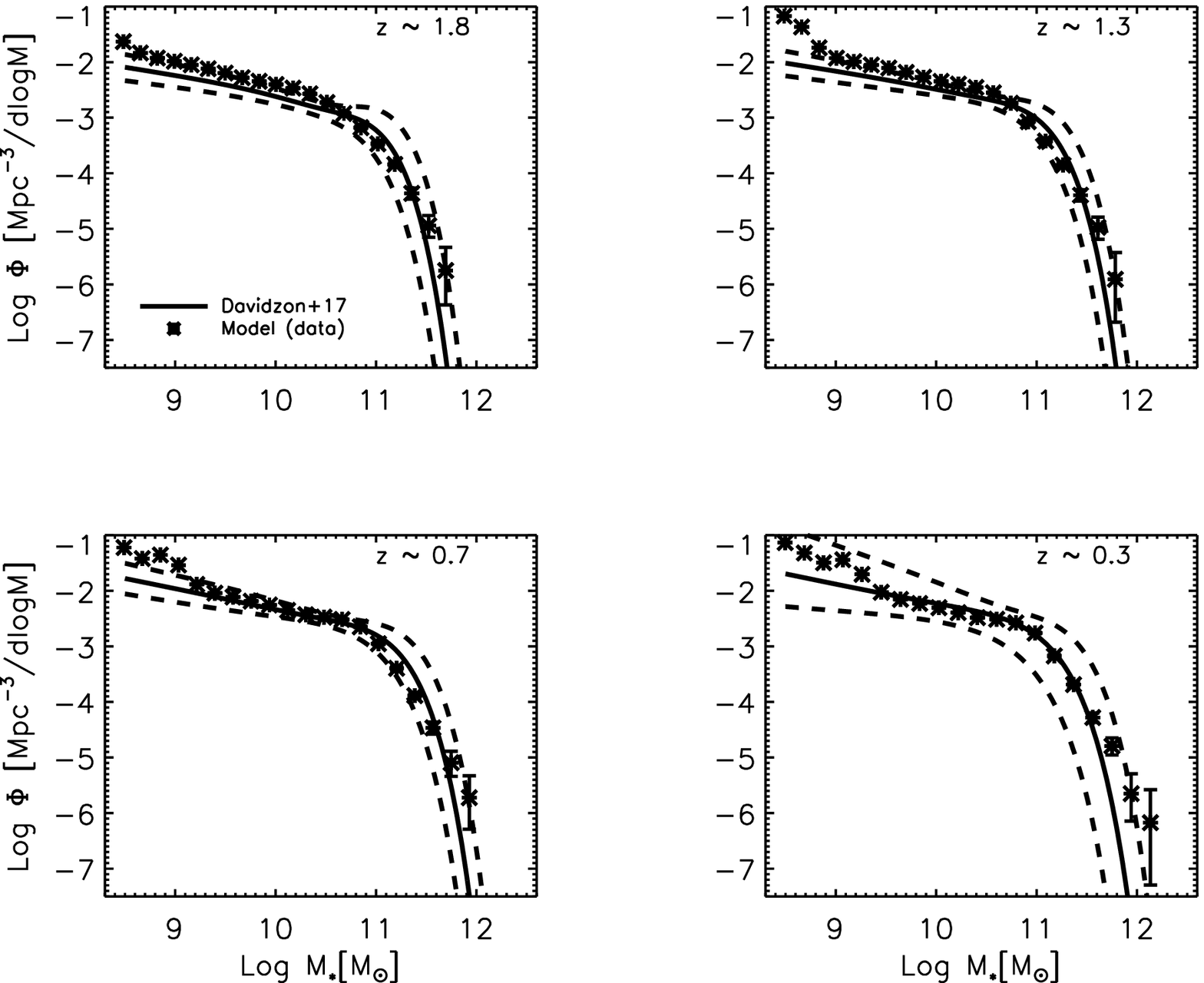} 
\caption{Evolution of the SMF from $z \sim 1.8$ down to $z \sim 0.3$ predicted by our model calibrated on the COSMOS survey at $z_{match}=2.3$. Stars represent model data,
while the dashed lines represent the \cite{davidzon17} fit of the COSMOS data.}
\label{fig:smf_cs}
\end{center}
\end{figure*}

\begin{figure*} 
\begin{center}
\includegraphics[scale=.9]{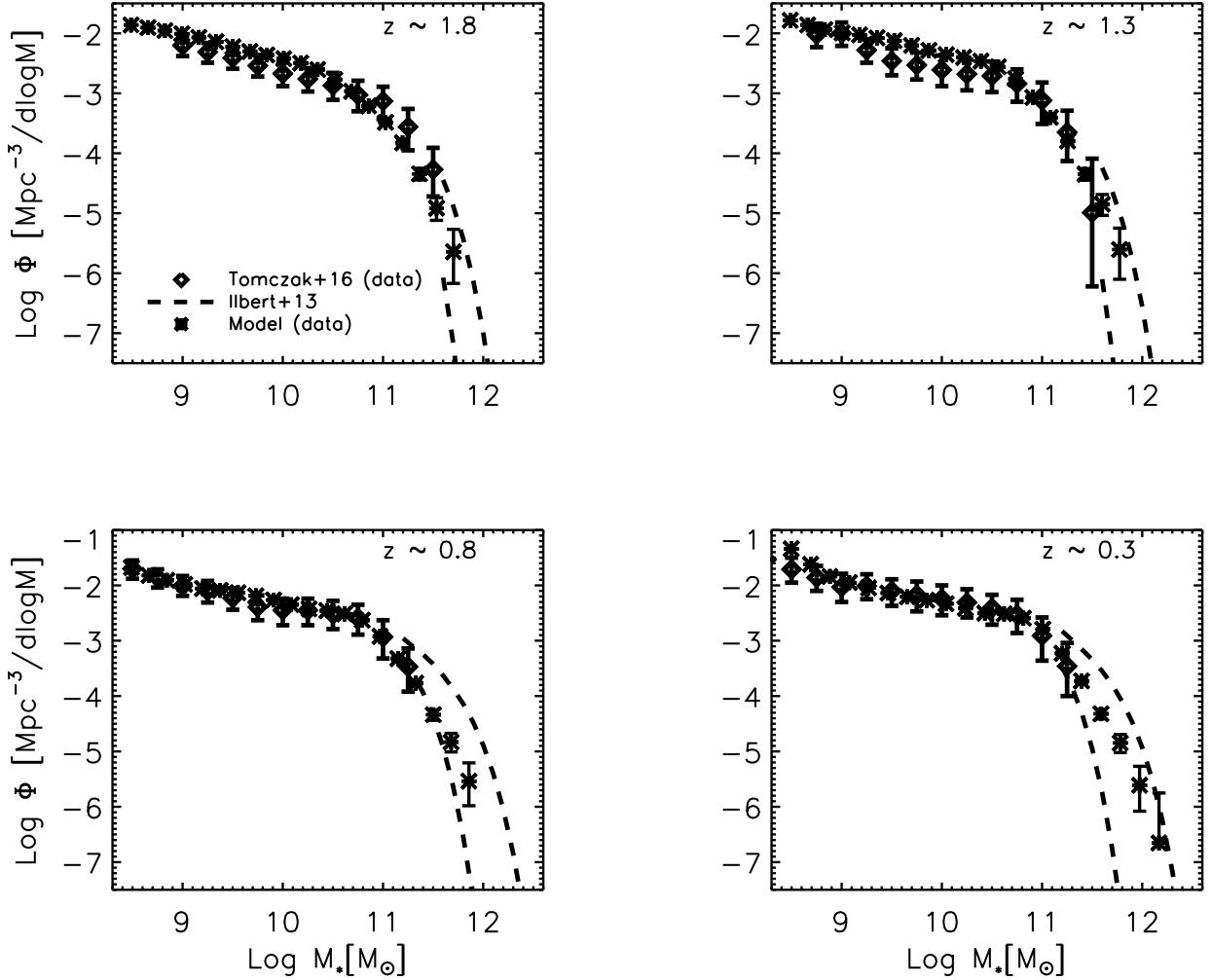} 
\caption{Evolution of the SMF from $z \sim 1.8$ down to $z \sim 0.3$ predicted by our model calibrated on our HYBRID survey (combination of ZFOURGE and UltraVISTA, as explained in the text) at $z_{match}=2.3$. 
Stars and diamonds represent model and observed data, respectively, while the dashed lines represent the \cite{ilbert13} fit of the UltraVISTA data in the high-mass end.}
\label{fig:smf_ib}
\end{center}
\end{figure*}

\begin{figure*} 
\begin{center}
\includegraphics[scale=.9]{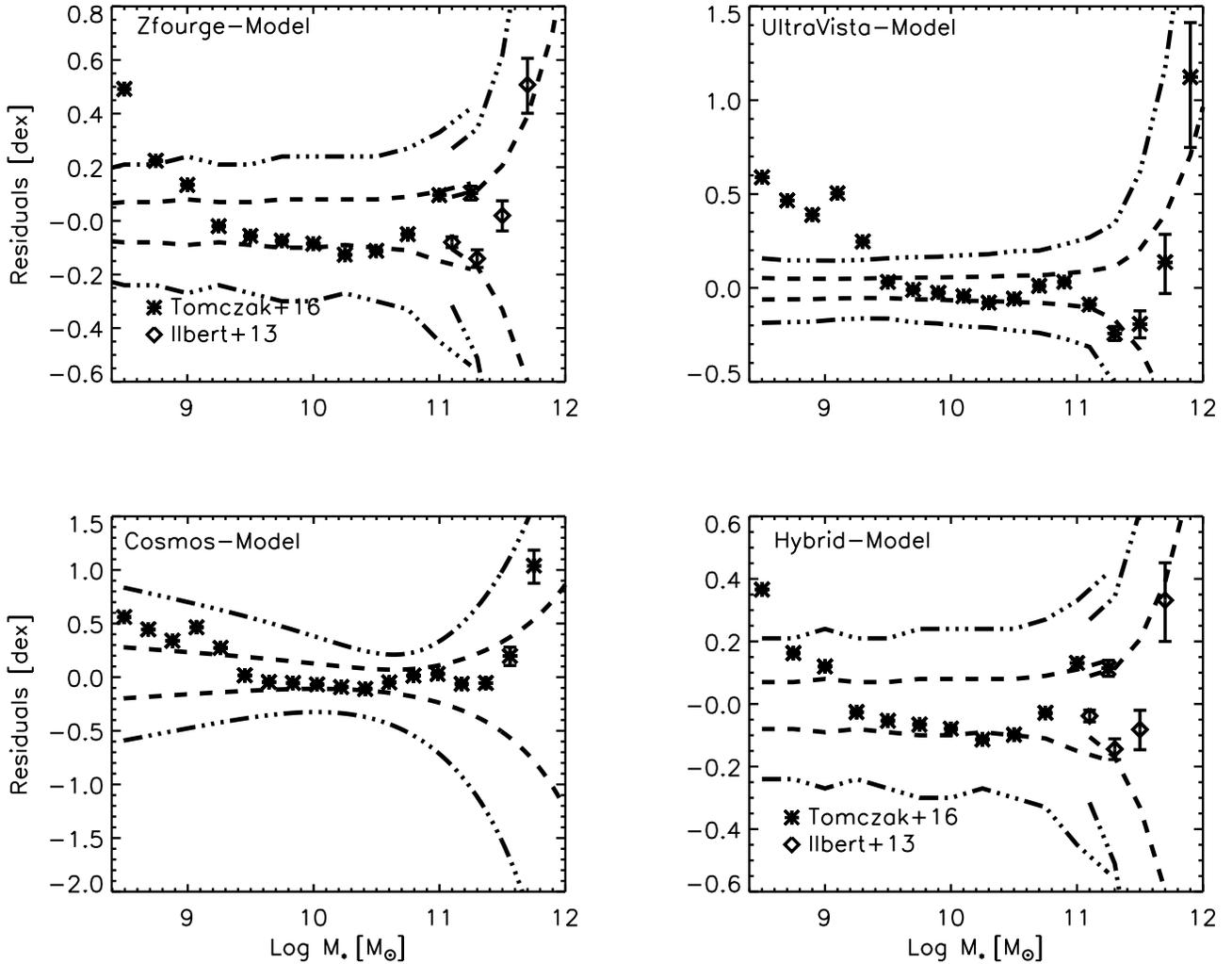} 
\caption{Residuals between the observed SMF and our model calibrated on them at $z \sim 0.3$. In each panel, stars (and diamonds in the top left and bottom right panels) represent the residuals, while with dashed and 
dash-dotted lines we plot the 1$\sigma$ and 3$\sigma$ scatter (observed), respectively.}
\label{fig:residuals}
\end{center}
\end{figure*}

\begin{figure*} 
\begin{center}
\includegraphics[scale=.9]{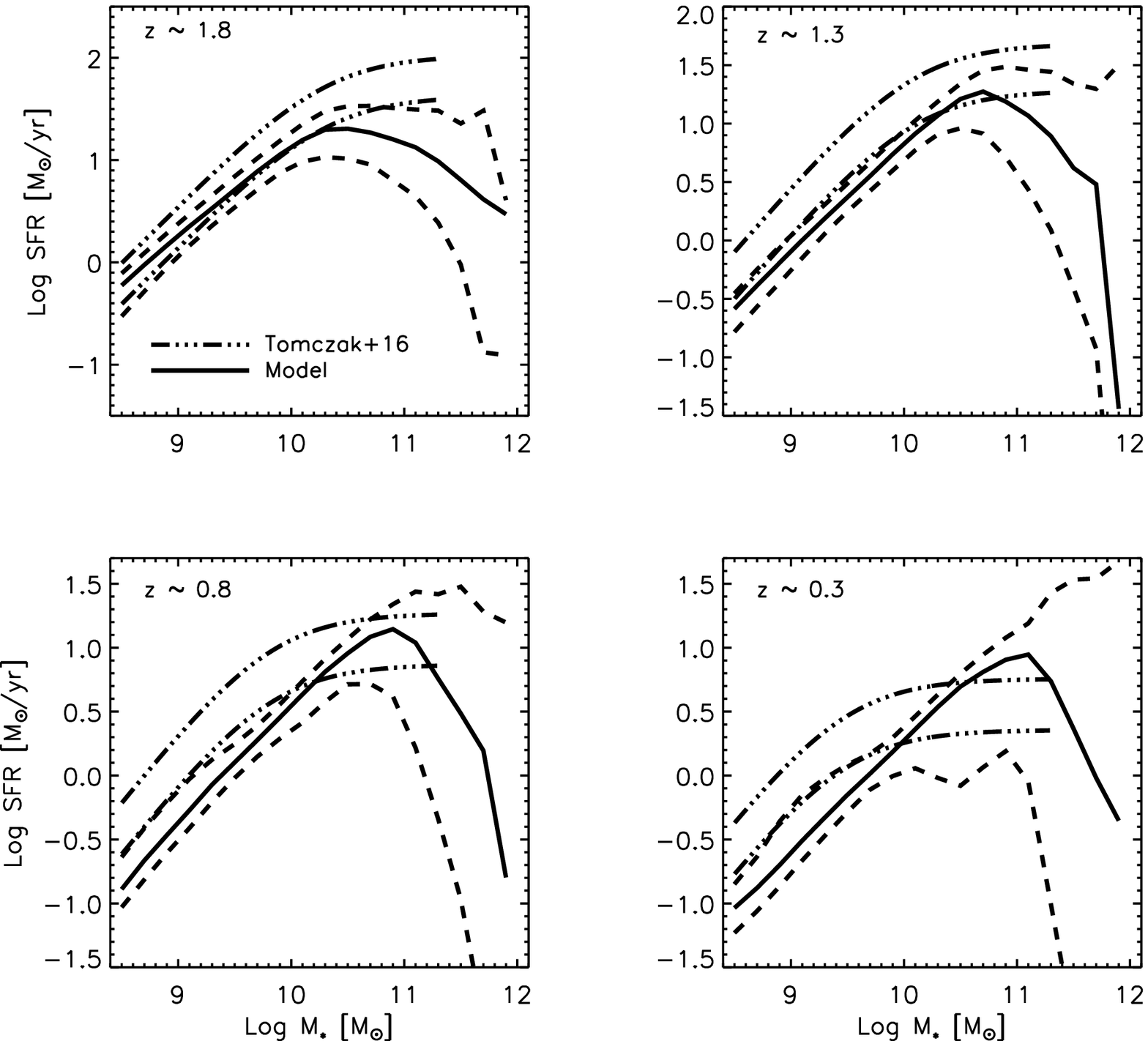} 
\caption{SFR$-M_*$ relation from $z \sim 1.8$ to $z \sim 0.3$. The solid lines represent the median SFR at each redshift, while the dashed lines show the 16th and 84th percentiles of the distribution. 
We compare our predictions with the 1$\sigma$ scatter region (dash-dotted lines) of the observed relation by \cite{tomczak16}.}
\label{fig:sfrmass}
\end{center}
\end{figure*}

In Figure \ref{fig:smf_zf} we show the evolution of the SMF predicted by the model and compared with the observed one, from $z\sim 1.8$ to $z\sim 0.3$. 
The model has been calibrated at $z_{match}=2.3$ using the observed SMF from ZFOURGE. Model and observed data are represented by stars and diamonds, respectively, while 
the dashed lines indicate the region between $\pm 3\sigma$ scatter observed by \cite{ilbert13} (UltraVISTA survey), which we add to the plot in order to have an idea 
of the shape of the SMF in the high-mass end (data from \citealt{tomczak14} do not go beyond $[10^{11.25}-10^{11.5}] M_{\odot}$). Overall, the model reproduces fairly well 
the evolution of the SMF down to $z\sim 0.3$. At $z\sim 1.8$ and in the stellar mass range $\sim [10^{9}-10^{10}] M_{\odot}$ the SMF is slightly overpredicted, while the 
trend changes in the knee and beyond. Considering the scatter, it is, however, possible to assert that the model reproduces the first gigayear of evolution. At $z\sim 1.3$ the 
prediction gets worse in the stellar mass range $\sim [10^{9.5}-10^{10.3}] M_{\odot}$, but still within the scatter, while the knee and the high-mass end are well 
reproduced. This means that there are not drastic deviations from the observed SMF after the second Gyr of evolution. At $z\sim 0.8$, the predicted SMF agrees very well 
with the observed one over the whole stellar mass range, and the trend persists down to $z\sim 0.3$, although the very low mass end is overpredicted and the very high mass 
end lies on the limit of the observed scatter. Hence, our model calibrated with ZFOURGE reproduces well about 8 Gyr of evolution of the SMF. 

In Figure \ref{fig:smf_uv} we show the predictions of our model calibrated with UltraVISTA. Model and observed data are represented by stars and solid lines, respectively, while 
the dashed lines indicate the region between $\pm 3\sigma$ scatter observed by \cite{ilbert13}. The evolution of the SMF appears different from that seen in 
Figure \ref{fig:smf_zf}, when the model is calibrated with ZFOURGE. At $z\sim 1.8$ and $z\sim 1.3$, the knee of the SMF predicted by the model is offset low from 
the observed one, while the low- and high-mass ends are within the observed scatter. From $z\sim 1.3$ to $z\sim 0.7$ ($\lesssim 3$ Gyr) there is a evolution of the low-mass 
end in the stellar mass range $[10^{8.5}-10^{9.2}] M_{\odot}$, which persists down to $z\sim 0.3$. At this redshift, the model is able to reproduce the SMF in the 
stellar mass range $[10^{9.5}-10^{11.9}] M_{\odot}$, but fails below $10^{9.5} M_{\odot}$, where the number density is substantially overpredicted. 

A similar picture can be drawn from Figure \ref{fig:smf_cs}, which shows the predictions of our model calibrated with COSMOS (symbols and lines have the same meaning as 
in Figure \ref{fig:smf_uv}) and a comparison with the \cite{davidzon17} data. Although the scatter in this case is distributed differently, the main features seen before still hold. 
At $z\sim 1.8$ and $z\sim 1.3$ the knee of the SMF predicted by the model is still offset low with respect to the observed one, although the mismatch is less evident if compared 
with the one found in Figure \ref{fig:smf_uv}. From $z\sim 1.3$ to $z\sim 0.7$, there is still the rapid evolution of the low-mass end in the stellar mass range 
$[10^{8.5}-10^{9.2}] M_{\odot}$ as seen in the previous case, which holds down to $z\sim 0.3$. If we compare figures \ref{fig:smf_uv} and \ref{fig:smf_cs} panel by panel, we see 
that the evolution of the SMF predicted by the model calibrated with UltraVISTA looks remarkably similar to the evolution of the SMF predicted by the model calibrated with 
COSMOS. This is interesting since, as we can see from Figure \ref{fig:surveys}, at $z=z_{match}$, COSMOS shows higher number densities in the low-mass end than 
UltraVISTA, and lower number densities beyond the knee. We will come back to this in Section \ref{sec:discussion}.

Figure \ref{fig:smf_ib} shows the SMF evolution as predicted by the model calibrated with our mock survey HYBRID, from $z\sim 1.8$ to $z\sim 0.3$ (symbols and lines have the 
same meaning as in Figure \ref{fig:smf_zf}). As explained in section \ref{sec:surveys}, HYBRID has been constructed by merging ZFOURGE and UltraVISTA data, in such a way to be 
sensitive to ZFOURGE data from the low-mass end to $10^{11} M_{\odot}$, and to UltraVISTA in the high-mass end. For this reason, the evolution of the SMF in the stellar mass
range $[10^{8.5}-10^{11.}] M_{\odot}$ is almost identical to the one predicted by the model calibrated with ZFOURGE, and the only appreciable difference arises in the high-mass 
end. In fact, at all redshifts shown, the high-mass end predicted by the model calibrated with HYBRID is closer to the average number density rather than to the upper limit, as 
predicted by the model calibrated with ZFOURGE. Overall, the two surveys provide very similar evolutions.

In order to better judge the goodness of each prediction, it is necessary to quantify the deviation of the model from the observed data. We address this point in Figure 
\ref{fig:residuals}, where we plot the residuals, that is, the difference between the logarithm of the observed number density and the logarithm of the predicted number density for 
the model calibrated with our surveys, at $z\sim 0.3$. In each panel, stars (and diamonds in the top left and bottom right panels) represent the residuals, while with dashed and 
dash-dotted lines we plot the 1$\sigma$ and 3$\sigma$ observed scatter, respectively. As commented above, ZFOURGE is very close to HYBRID and UltraVISTA to COSMOS. What appears to 
be very interesting is the level of accuracy of the predictions after about 8 Gyr of evolution of the SMF. It is worth discussing them one by one. In the top left panel of Figure 
\ref{fig:residuals}, we show the residuals between ZFOURGE and the model. In the stellar mass bin $[10^{9.3}-10^{11.5}] M_{\odot}$, residuals lie within $1\sigma$ observed 
scatter, and most of them are very close to zero. If we consider $3\sigma$ (observed scatter), only the first two data points lie outside it. We calculated the mean residual over 
the stellar mass range shown, and the same quantity not considering the two largest residuals (which are always the first and last data points in all panels). In the case of 
ZFOURGE, the mean residual is 0.15 dex, which reduces to 0.09 dex if we do not consider the first and last data points (see Table \ref{tab:residuals}).

Residuals remarkably increase in the top right (UltraVISTA) and bottom left (COSMOS) panels. Here the model reproduces very well the SMF in the stellar mass range $[10^{9.5}-10^{11.5}] M_{\odot}$,
but residuals are very large outside this range of mass. Indeed, as reported in Table \ref{tab:residuals}, the mean residuals (both cases) are much larger than those discussed 
above (ZFOURGE). As one can expect, in the case of HYBRID, residuals are very close to those shown in the top left panel (ZFOURGE) and somewhat lower for the first two and 
last data points. In the case of HYBRID, the mean residual is 0.12 dex, which reduces to 0.09 dex if we do not consider the first and last data points (see Table \ref{tab:residuals}).

These results are remarkably important considering that no study so far, focused on this topic and following a similar approach (e.g., \citealt{leja15,tomczak16,contini17}), 
has found an average residual lower than $\sim 0.3$ dex. In the light of this, in the next section we will fully discuss these results and their importance to the goals of this study.

\begin{table}
\caption{Mean residuals (in dex) at $z \sim 0.3$ between the observed and model SMFs, considering all data points (first line), and the same but excluding the two most deviant points} (second line).
\begin{center}
\begin{tabular}{llllll}
\hline
Survey & HYBRID & ZFOURGE & UltraVISTA & COSMOS \\
  --   & 0.12  & 0.15    &  0.24     &  0.22  \\
  --   & 0.09  & 0.09    &  0.16     &  0.14  \\
\hline

\hline
\end{tabular}
\end{center}
\label{tab:residuals}
\end{table}

\section{Discussion}
\label{sec:discussion}

As discussed in Section \ref{sec:intro}, many attempts have been made in recent years to reconcile the evolution of the SMF with the observed $SFR-M_*$
relation. The large amount of effort spent on this topic by several authors resulted in the conclusion that the slope of the $SFR-M_*$ relation must be mass and redshift 
dependent. Moreover, this relation alone cannot explain the evolution of the SMF, but other physical processes responsible for galaxy growth, such as mergers and 
stellar stripping, must be taken into account. Nevertheless, once any possible process is properly considered, there is still a nonnegligible mismatch between the 
observed SMF and the predicted one. Understanding the reason behind that is the main goal of this study.

Recently, \cite{steinhard17}, by means of an analytic model, focused on the low-mass end of the SMF and confirmed that the growth of galaxies along the main sequence 
alone cannot reproduce the observed SMF. As a possible solution, they suggest that mergers are also necessary to describe the growth of galaxies in the stellar mass range
they investigated. This is qualitatively in good agreement with \cite{tomczak16}, but not quantitatively. In fact, \cite{tomczak16} show that mergers can help to lower 
the discrepancy between the observed and inferred SMFs if low-mass galaxies merge with more massive galaxies, but the rate required would imply that between 25\% and 65\%
of low-mass galaxies have to merge with a more massive galaxy per gigayear. This rate appears to be too high if compared with current estimates of galaxy merger rates (e.g. 
\citealt{lotz11,williams11,leja15}).
In PapI, besides mergers, we considered stellar stripping. We showed that mergers and stellar stripping have opposite effects, and a full modeling alleviates the 
discrepancies between the observed and theoretical SMF, but not enough for a full match (within the observed scatter) along the whole stellar mass range investigated.  

Several arguments have been invoked in order to explain the residuals between the two SMFs, which appear to be a function of redshift and stellar mass. The most credible 
ones rely on the accuracy in the measurements of both stellar mass and SFR (\citealt{leja15,tomczak16,contini17}). In fact, in PapI we showed that once we reduce $z_{match}$ 
we obtain a better agreement at low redshift, which means that either measurements of stellar mass and SFR at lower redshift are less affected by uncertainties, or the 
scatter along these measurements has less time to propagate as time passes. \cite{tomczak16} let the SMF evolve for a limited amount of time and for different redshift 
bins, finding residuals similar to what we find in PapI. They conclude that either the SFR measurements are overestimated or the growth of the \cite{tomczak14} 
mass function is too slow, or both.

One of the goals of this study is to considerably reduce the residuals between the observed and inferred SMFs. In order to do it, we slightly changed the approach followed 
in PapI, and rather than assigning SFRs according to the $SFR-M_*$ relation at each time step, the model builds SFHs that depend on the galaxy type, stellar mass, and quenching 
timescale. As shown in Section \ref{sec:results}, the same model calibrated with different surveys leads to different evolutions of the SMF. This implies that there might 
be an intrinsic inconsistency in the growth of the SMF described by the same survey. As pointed out in PapI, the inferred evolution of the SMF is sensitive 
to the shape of the observed SMF at $z=z_{match}$. The differences in the evolution of the SMF from our four surveys are mainly the consequence of the gap between them 
in the low-mass end at $z=z_{match}$ (left panel of Figure \ref{fig:surveys}). COSMOS and UltraVISTA show at $z \sim 0.3$ the largest residuals in the low-mass end, 
because they are also higher than ZFOURGE and HYBRID in the very low mass range at $z=z_{match}$ \footnote{It must be noted that our model intrinsically predicts an higher number 
of low-mass galaxies, regardless of the survey chosen for calibration. This issue is discussed in the Appendix.}. 

The calibration of the model with the observed SMF at $z=z_{match}$ is not trivial. High number densities in the low-mass range will end up in too many galaxies
at low redshift. The calibration is then very sensitive to both the slope $\alpha^*$ and the normalization $\phi^*$ of the SMF. The observed SMFs agree at $z \sim 0.3$ and 
those predicted by the model show minimal residuals in most of the stellar mass range at the same redshift. This means that the small (but appreciable) offset seen at 
$z=z_{match}$ in the low-mass range affects the evolution of the modeled SMFs in the low-mass range only, while the offset in the rest of the stellar mass range can explain
the different residuals among the modeled SMFs in the high-mass end. To test the importance of the calibration of the model, we introduced a mock survey called HYBRID, which is,
as explained in Section \ref{sec:surveys}, a combination between ZFOURGE (sensitive to the low-mass end and intermediate mass) and UltraVISTA (sensitive to the high-mass end). 
The evolution of the SMF predicted by the model calibrated with HYBRID is very consistent with the evolution of the SMF predicted by the model calibrated with ZFOURGE. This 
strengthens the above argument that a minimal offset in the low-mass end at $z=z_{match}$ can remarkably affect the evolution of the SMF down to low redshift in the same 
stellar mass range. 

ZFOURGE and HYBRID show the smallest average residuals at $z \sim 0.3$, which are one-half of the ones shown by COSMOS and UltraVISTA (see Table \ref{tab:residuals}). This is completely 
due to the residuals in the low-mass end (see Figure \ref{fig:residuals}). In fact, if we only consider the stellar mass range $[10^{9.4}-10^{11.5}] M_{\odot}$, the mean residuals 
are very similar and between 0.04 and 0.06 dex for all the surveys. In a stellar mass range spanning over two orders of magnitude, our model predicts an SMF at $z \sim 0.3$ very close to the 
observed one and within $1\sigma$ observed scatter. If we consider the whole stellar mass range investigated, $\sim [10^{8.5}-10^{11.9}] M_{\odot}$, our average residuals are still below 
any other found in the past studies and, especially in the case of ZFOURGE (as well as HYBRID), are particularly small ($\leq 0.15$ dex). The improvement with respect to the results found 
in PapI is tangible (see Fig. 5 of PapI), where, in the case of ZFOURGE, residuals lie within the observed scatter in the stellar mass range $\sim [10^{10}-10^{11.3}] M_{\odot}$.

Our new model can make a prediction of the evolution of the $SFR-M_*$ relation as a function of time, since the SFR is only initialized at 
$z_{match}$ or $z_{form}$, and the SFH of each galaxy will depend on the time spent as a central or satellite and therefore on its quenching timescales $\tau_c$ and $\tau_s$. We show the 
prediction of our model in Figure \ref{fig:sfrmass}, from $z\sim 1.8$ to $z\sim 0.3$. The solid lines represent the median SFR at each redshift, and the dashed lines show the 16th and 84th 
percentiles of the distribution, while the dash-dotted lines indicate the 1$\sigma$ scatter of the observed relation by \cite{tomczak16} that we use for comparison. What appears to be 
relevant is the slope of the $SFR-M_*$ relation predicted by the model, which is consistent with the observed one at any redshift up to intermediate-mass galaxies. The SFR reaches a peak 
and then drops in the stellar mass range of massive galaxies, and the peak moves toward higher mass as the redshift decreases. This is because galaxies grow in mass (move to
the right of the plot), and a large part of massive galaxies are centrals (slower quenching). Another interesting feature concerns the comparison with the observed relation. Our predictions 
are indeed biased low with respect to the observed ones at every redshift, which implies that the model on average requires a lower SFH for each galaxy to fairly match the SMF at low redshift.

The features shown in Figure \ref{fig:sfrmass} are independent of the calibration (no dependence on the survey), and this has an important consequence. In fact, assuming that the observed 
$SFR-M_*$ relation at $z\sim 0.3$ is well constrained, the model would require a higher normalization of the relation at any redshift, that is, also at $z_{match}$. This means that, in order to 
match the SMF at low redshift, either the quenching timescales must be somehow shorter in order to recover the $SFR-M_*$ relation at any redshift, or the SFRs at $z_{match}$ are underestimated, 
or a combination of the two. By looking at the evolution of the SMF (independently of the survey), SFRs down to $z\sim 1.3$ do not seem to be underestimated since the predicted SMF lies 
above the observed one from the low-mass end to stellar masses around $10^{10.5} M_{\odot}$. In these redshift and stellar mass ranges, the model would require either shorter quenching timescales 
or lower SFRs. Below $z\sim 1.3$, the evolution of the SMF is consistent with our modeling of the quenching timescales, which in turn brings to lower SFRs than observed. In PapI we 
show that the combination of the SMF and $SFR-M_*$ relation measured by ZFOURGE leads to overpredicting, on average, the number densities at low redshift. As found by \cite{tomczak16}, who took 
advantage of the same survey, there is an inconsistency between stellar mass and SFR measurements, which agrees with our results in PapI. 

In the light of the results found in this study, and considering the parameter space used by the model, it is not possible to \emph{a priori} state which of the two relevant measurements,
stellar mass or SFR, is not consistent with the global evolution of the SMF. However, if the observed SMF at $z_{match}$ is correct, the $SFR-M_*$ relation must have a lower normalization at 
any redshift investigated in order to reproduce the observed SMF at low redshift. On the other hand, if we trust the relation between SFR and stellar mass as a function of redshift, the observed 
evolution of the SMF has to be faster. Considering that we used three different surveys (and a mock combination of two of them) and the above picture holds for each survey, it is plausible to 
support the first scenario, that is, the observed SFRs are overestimated (see, e.g., \citealt{wilkins08,kang10}).

For the sake of clarity, we must also note that our modeling does not explicitly consider any prescription for the mass quenching as described in \cite{peng10}. In that work, mass quenching 
is intended as the quenching of star forming galaxies around and above $\phi^*$ that follows a rate statistically proportional to the star formation rate (see their equation 17). Their model 
aims to consider the quenching given by internal feedback, such as supernova and active galactic nucleus feedback. Despite the goal being the same, in this work we consider a mass-dependent quenching, where the time 
scale for quenching is inversely proportional to the galaxy stellar mass (faster quenching for more massive galaxies). These two definitions are conceptually different and likely the reason why 
(or at least in part) we overpredict the number density of very massive galaxies, but both aim to consider the internal feedback that quenches galaxies. Moreover, we must note that,
on one hand, our analytic definitions for the quenching timescales are theoretically valid, but on the other hand, they are also quite arbitrary in nature.

\section{Conclusions}
\label{sec:conclusions}

We have studied the evolution of the SMF from $z \sim 2.3$ to $z \sim 0.3$ as predicted by an analytic model calibrated with observed data. We set the model in order 
to perfectly match the observed SMF at $z_{match}$, assign an SFR to galaxies by means of the SFR-$M_*$ relation at $z_{match}$ or $z_{form}$, and let the SFRs evolve according to 
functional forms that mainly depend on two characteristic quenching timescales. We took advantage of observed SMFs from four surveys: ZFOURGE (\citealt{tomczak14}), UltraVISTA 
(\citealt{ilbert13}), COSMOS (\citealt{davidzon17}), and a mock combination of ZFOURGE and UltraVISTA that we have called HYBRID. From the predictions of our model, we can conclude 
the following:
\begin{itemize}
 \item The inferred evolution of the SMF is in good agreement with the observed one, and the level of accuracy of our modeling mostly depends on the survey used for the calibration. 
       We confirm the result found in PapI: the same model calibrated with different surveys leads to different evolutions of the SMF, which are very sensitive to the shape of 
       the observed SMF at $z=z_{match}$, and in particular to the low-mass end at $z=z_{match}$. Although the four observed SMFs at $z=z_{match}$ look different, they evolve to the same 
       observed SMF at $z \sim 0.3$, which implies an intrinsic inconsistency in the growth of the SMF described by the same survey.
       
 \item Our new model brings to a much better agreement between the observed and inferred SMFs at $z \sim 0.3$ with respect to previous results. In fact, the residuals between the two SMFs 
       are located within $1\sigma$ observed scatter in most of the stellar mass range investigated, and within $3\sigma$ in almost the whole range. HYBRID and ZFOURGE are those which 
       provide the smallest mean residuals, around 0.12/0.15 dex, which reduce to 0.09/0.09 dex if we do not consider the very first and last data points.
      
 \item The $SFR-M_*$ relation predicted by our model is offset low with respect to the observed one (\citealt{tomczak16}) at any redshift and independent of the calibration at $z=z_{match}$.
       \cite{tomczak16} find that either the evolution of the observed SMF is too slow or the SFR measurements are overestimated. As discussed in Section \ref{sec:discussion}, the latter is 
       more plausible since the prediction of the $SFR-M_*$ relation does not depend on the survey used to calibrate the model.
\end{itemize}
Considering the fact that our model produces fairly well about 8 Gyr of evolution of the SMF (with average residuals between 0.04 and 0.06 dex in the stellar mass range 
$[10^{9.4}-10^{11.5}] M_{\odot}$), the improvement over PapI and previous studies is remarkable. However, in order to investigate if the relation between the quenching timescales 
and stellar mass (with redshift) predicted by the model is quantitatively and qualitatively correct, in a forthcoming paper we aim to apply the model to reproduce the evolution 
of the SMF of star-forming and quiescent galaxies separately. We will make use of a color separation in order to split our sample into the two subpopulations of galaxies, and we
will test our model predictions against several observed quantities, such as the fraction of quiescent galaxies as a function of stellar mass and redshift.

\section*{Acknowledgements}
E.C. acknowledges financial support from the Chinese Academy of Sciences Presidents' International 
Fellowship Initiative, Grant No.2015PM054.
E.C., X.K., and A.R. acknowledge financial support by the 973 program (No. 2015CB857003, 2013CB834900),
NSF of Jiangsu Province (No. BK20140050), and the NSFC (No. 11333008,11550110182,11550110183). 
E.C. and S.K.Y. acknowledge support from the Korean National Research Foundation 
(NRF-2017R1A2A1A05001116) and from the Brain Korea 21 Plus Program (21A20131500002). This study was 
performed under the umbrella of the joint collaboration between Yonsei University Observatory 
and the Korean Astronomy and Space Science Institute.

\label{lastpage}

\appendix

\section{Testing the low-mass end}
\label{sec:append}

\begin{figure*} 
\begin{center}
\includegraphics[scale=.9]{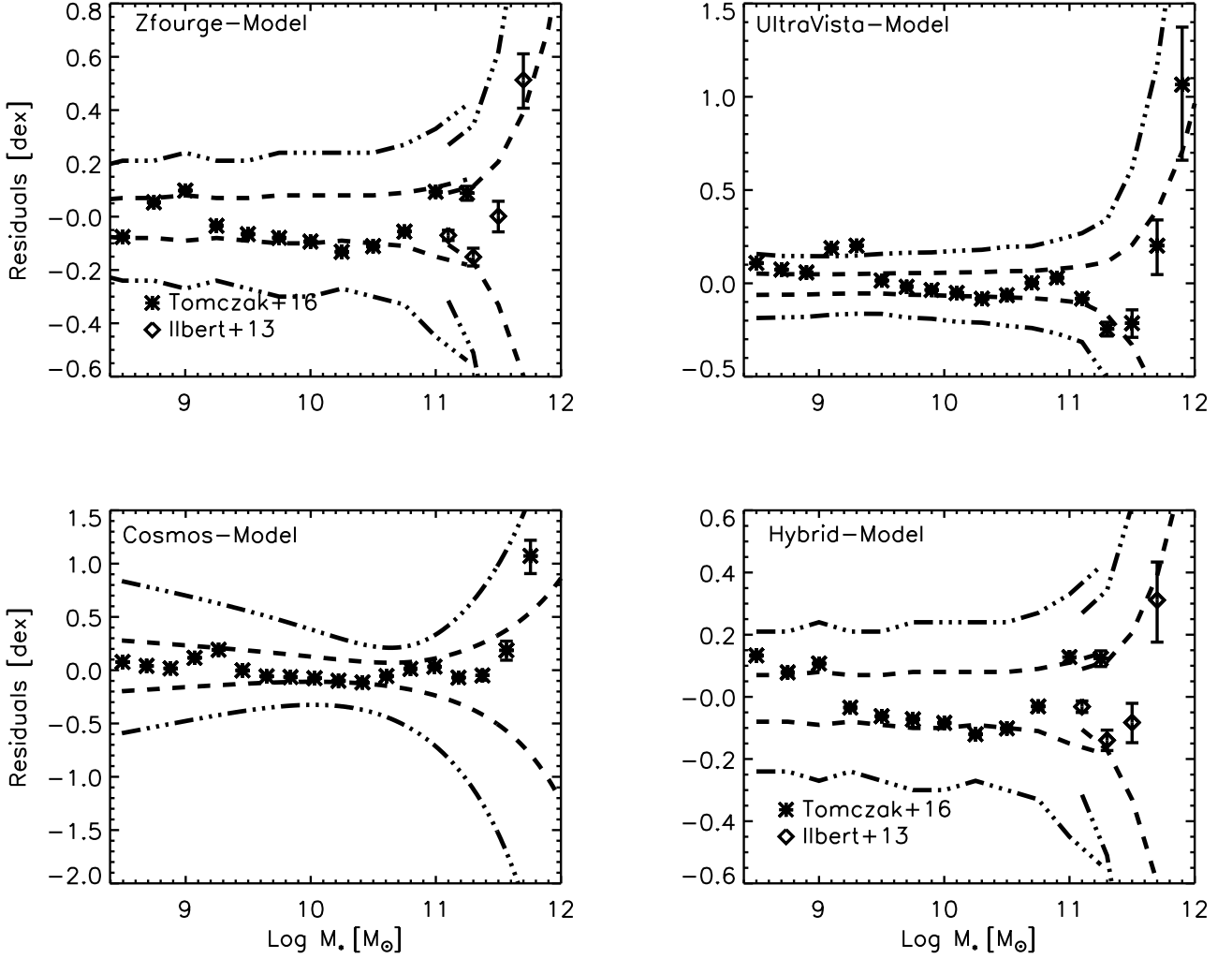} 
\caption{Residuals between the observed SMF and our model calibrated on them at $z \sim 0.3$, obtained by quenching low-mass galaxies below a given mass cut (different for different surveys, see text). 
In each panel, stars (and diamonds in the top left and bottom right panels) represent the residuals, while with dashed and dash-dotted lines we plot the 1$\sigma$ and 3$\sigma$ scatter (observed), respectively.}
\label{fig:residuals_lmt}
\end{center}
\end{figure*}

The low-mass end of the SMF, below approximately $\log M_* [M_{\odot}] \sim 9.3-9.4$ depending on the survey used to calibrate the model, is not well constrained. The residuals in 
this stellar mass range are perceptibly higher than the others, with the exception of the very high mass end, at low redshift. In principle, this can be a consequence of the fact that 
the slope of the observed SMF at $z \sim 2.3$ that we use to calibrate the model is too steep. However, for UltraVISTA and COSMOS, as we can see in Figs \ref{fig:smf_uv} and 
\ref{fig:smf_cs}, the low-mass end starts to diverge already between $z \sim 1.8$ and $z \sim 1.3$, while this is not seen in the case of ZFOURGE and HYBRID, where the low-mass end 
starts to diverge at $z \sim 0.3$. Another additional explanation for the divergence of the low-mass end comes directly from the merger trees and the way the model populates subhaloes 
with galaxies. In fact, it is well known that merger trees extracted from N-body simulations usually predict too many low-mass subhaloes that survive down to the present day (see, e.g., 
\citealt{lapi17} and references therein). A rough way to test whether it is possible to find a better agreement between the observed and predicted SMFs at low redshift in the low-mass 
end would imply the suppression or quenching of newborn galaxies residing in such subhaloes. In order to understand the limit of our modeling, we run the four flavors of the model calibrated 
with our surveys by adding a stellar mass cut below which galaxies that form after $z=z_{match}$ are not allowed to form stars anymore and quench instantly.  
Figure \ref{fig:residuals_lmt} shows the residuals between the observed and predicted SMFs for all models as a function of stellar mass, at $z \sim 0.3$. The cut in stellar mass has been 
chosen (for each model) such that data points below $\log M_* [M_{\odot}]=9$ in Figure \ref{fig:residuals_lmt} stayed within $2\sigma$ observed scatter. Not surprisingly, we find that the 
model calibrated with different surveys requires different cuts. HYBRID is the one that requires the lowest cut, $\log M_{cut} [M_{\odot}]<7.75$, while the other three require 
$\log M_{cut} [M_{\odot}]<8.15$ (ZFOURGE) and $\log M_{cut} [M_{\odot}]<8.20$ (for both UltraVISTA and COSMOS). Moreover, these cuts bring the mean residual (not considering the very last
data point of the high-mass end) below 0.1 dex for all models and a much better agreement with observations in particular for the model calibrated with UltraVISTA and COSMOS, which are 
the surveys that show the highest number densities below $\log M_* [M_{\odot}]=8.5$ at $z=z_{match}$. This is also the reason why the cut in mass is higher in these two cases. 
To conclude, we must note that such cuts have a negligible effect in the rest of the stellar mass range.

\end{document}